\documentstyle[aps]{revtex}
\begin{document}
\draft
\title{
Note on the Spin-Spin Relaxation Time 
in High-$T_{\rm c}$ Cuprate Superconductors
\cite{N}
}
\author{
O. Narikiyo
}
\address{
Department of Physics, 
Kyushu University, 
Fukuoka 810-8560, 
Japan
}
\date{
November, 2000
}
\maketitle
\begin{abstract}
Recent experiments on the spin-spin relaxation time 
in high-$T_{\rm c}$ cuprate superconductors 
are discussed on the basis of the BCS-RPA theory. 
\vskip 8pt
\noindent
{\it Keywords:} spin-spin relaxation time, 
high-$T_{\rm c}$ cuprate superconductor
\end{abstract}
\vskip 18pt
  The temperature dependence of 
the spin-spin relaxation time recently measured 
in high-$T_{\rm c}$ cuprate superconductors
\cite{Itoh1,Itoh2,Itoh3,Tokunaga} 
deviates from the expected behavior of a prevailing thoery.\cite{BS1} 
  The observed relaxation time 
has strong temperature dependence 
for temperatures below the superconducting transition temparature, 
$ T < T_{\rm c} $, 
while the theory predicts little temparature dependence. 
  Such a temperature independence predicted by the theory 
was regarded as an evidence 
for the $d$-wave superconductivity 
and established by the measurements in La- and Y-systems.\cite{Itoh4} 
  On the other hand, the measurements in Tl-, Hg- and Bi-systems
\cite{Itoh1,Itoh2,Itoh3,Tokunaga} 
revealed the insufficiency of the theory.\cite{BS1} 

  In this Short Note 
we clarify that the temperature independence 
is not a direct consequence of the theory 
by searching much wider parameter space than in the original work\cite{BS1} 
and that the recent experiments\cite{Itoh1,Itoh2,Itoh3,Tokunaga} 
can be understood 
within the framework of the theory. 

  The Gaussian component of the spin-spin relaxation time 
$T_{\rm G}$ is expressed as 
\begin{equation}
{T_G(T_c) \over T_G(T)}=\left[ {R(T) \over R(T_c)} \right]^{1/2},
\end{equation}
using 
\begin{equation}
R(T)=\sum_{\bf q} F^4({\bf q})\chi^2({\bf q})
    -\left[ \sum_{\bf q} F^2({\bf q})\chi({\bf q}) \right]^2.
\end{equation}
  Here $\chi({\bf q})$ is the static spin susceptibity 
and $F({\bf q})$ is the form factor. 

  The static spin susceptibity is given by the BCS-RPA theory\cite{BS1} as 
\begin{equation}
\chi({\bf q})={\chi_0({\bf q}) \over 1 - U \chi_0({\bf q})}, 
\end{equation}
with 
\begin{equation}
\chi_0({\bf q})={1 \over N}\sum_{\bf q}
\left[ C_+({\bf p},{\bf q})
       {f(E_{\bf p+q})-f(E_{\bf p}) \over E_{\bf p}-E_{\bf p+q}} 
     + C_-({\bf p},{\bf q})
       {f(-E_{\bf p+q})-f(E_{\bf p}) \over E_{\bf p}+E_{\bf p+q}}
\right],
\end{equation}
where $U$ is the parameter for the on-site Coulomb interaction 
and $f(E)$ is the Fermi distribution function. 
  The energy for the BCS quasiparticle is given by 
\begin{equation}
E_{\bf p}=\sqrt{\varepsilon_{\bf p}^2+\Delta_{\bf p}^2},
\end{equation}
with 
\begin{equation}
\varepsilon_{\bf p}=-2t[\cos(p_x a)+\cos(p_y a)]
                    -4t'\cos(p_x a)\cos(p_y a)
                    -2t''[\cos(2p_x a)+\cos(2p_y a)]
                    -\mu,
\end{equation}
where $t$, $t'$, $t''$ is the hopping parameter 
and $\mu$ is the chemical potential 
and 
\begin{equation}
\Delta_{\bf p}={\Delta_0(T) \over 2}[\cos(p_x a)-\cos(p_y a)],
\end{equation}
where $\Delta_0(T)$ is the gap function in the BCS theory. 
  Here we consider a tight-binding model on the square lattice 
of the lattice constant $a$. 
  The coherence factor is given by 
\begin{equation}
C_\pm({\bf p},{\bf q})={1 \over 2}
\left[ 1 \pm {\varepsilon_{\bf p+q}\varepsilon_{\bf p}
              +\Delta_{\bf p+q}\Delta_{\bf p} \over
              E_{\bf p+q}E_{\bf p} } \right].
\end{equation}
  The form factor is given by 
\begin{equation}
F({\bf q})=A_{cc}+2B[\cos(q_x a)+\cos(q_y a)], 
\end{equation}
in the present two-dimensional tight-binding model. 

  We have done numerical calculations with above formulae. 
  The parameters are fixed as $t'/t=-1/6$, $t''/t=1/5$, $\mu/t=-1$, 
$T_{\rm c}/t=0.1$ and $U/t=3$. 
  The temperature dependence of the chemical potential $\mu$
has been neglected. 
  The choice of $T_{\rm c}/t=0.1$ is in accordance with ref.\ 5 
and $U/t=3$ is consistent with ref.\ 7. 
  Our main results have been checked to be insensitive 
to the choice of the values of $t'$ and $t''$. 

  First, in Fig.\ 1(a), we have examined the temperature dependence 
of the static spin susceptibility. 
  The susceptibility is strongly temperature dependent 
around ${\bf q}=(0,0)$ and weakly dependent around ${\bf q}=(\pi/a,\pi/a)$. 
  The latter reflects the fact 
that the existence of the $d$-wave superconductivity 
does not conflict with the antiferromagnetic long-range correlation. 

  Since the form factor is fixed as $A_{cc}=-4B$ in ref.\ 5, 
the temperature dependence of $T_{\rm G}$ is predominantly 
determined by that of $\chi({\bf q})$ around ${\bf q}=(\pi/a,\pi/a)$. 
  Namely, the contribution from the region around ${\bf q}=(0,0)$ 
is masked by the form factor 
so that $T_{\rm G}$ has little temperature dependence. 
  Such a chice is consistent with the data for La- and Y-systems. 
  For example, $A_{cc}/B=-177/44$ in YBa$_2$Cu$_4$O$_8$.\cite{Itoh6} 

  On the other hand, 
the masking is absent in other cuparate superconductors. 
  For example, $A_{cc}/B=-170/100$ in Tl$_2$Ba$_2$CuO$_6$\cite{Itoh6} 
and the numerical result for $T_{\rm G}$ with this value in Fig.\ 1(c) 
has substantial temperature dependence. 
  The 15\% reduction at the lowest temparature 
is twice of the reduction in the theory of ref.\ 5 
and consistent with the experiment.\cite{Itoh1} 

  It can be said that we have substantial temperature dependence 
if the masking is absent. 
  However, the temperature dependence observed 
in Hg- and Bi-systems\cite{Itoh2,Itoh3,Tokunaga} 
is much stronger than obtained in Fig.\ 1(c). 
  As seen in Fig.\ 1(b) where $A_{cc}/B=-170/100$, 
the shorter the coherence length 
of the superconductivity, the stronger the temperature dependence is. 
  Although more elaborate treatment is needed for 
the short coherence-length (strong coupling) case, 
the qualitative feature can be captured 
by the present parameter extrapolation. 
  In ref.\ 5 the parameter has been fixed as 
$2\Delta_0(0)/k_{\rm B}T{\rm c}=3.52$ of the weak coupling value. 
  Thus it can be said that we have a strong temperature dependence 
if the coherence length is short. 
  Actually in the numerical result 
with parameters appropriate for Bi-system of Fig.\ 1(d), 
we have a strong temparature dependence as commented in ref.\ 4. 

  In summary, the temperature dependence of the spin-spin 
relaxation time observed by recent experiments can be understood 
within the BCS-RPA theory, 
if the coherence length of the superconductivity is short and 
the masking due to the form factor is absent. 
  Such a conclusion is easily understood 
by taking account of the fact\cite{BS2} 
that the BCS susceptibility $\chi_0({\bf q})$ 
is suppressed by the Cooper pair (singlet) formation 
for $|{\bf q}|<\xi^{-1}$ 
where $\xi$ is the coherence length for the superconductivity 
(the size of the Cooper pair). 

  Since our calculation is based on the mean-field approximation, 
it can not directly fit experimantal data and 
will be modified if fluctuations are taken into account. 
  The superconducting transition tempetatute $T_{\rm c}$ 
in the present theory corresponds to the temperature 
giving the maximum of $1/T_{\rm G}$ in experiments 
and is higher than experimentally observed $T_{\rm c}$. 
  The steep decrease of $1/T_{\rm G}$ around $T=T_{\rm c}$ 
in the theory will become gentle due to fluctuations. 

  As noted in ref.\ 4, 
the on-set temperature for the decrease of $1/T_{\rm G}$ 
due to superconducting fluctuations in experiments 
is lower than the so-called pseudogap temperature 
observed by the spin-lattice relaxation time of the NMR 
and the ARPES experiments. 
  The on-set temperature seems to correspond to the temperature 
giving the maximum of the Hall coefficient\cite{Ong1,Ong2} 
and this maximum temperature can be interpreted as the on-set 
temperature for superconducting fluctuations, 
since the collective contribution of superconducting fluctuations 
to the Hall coefficient 
has opposite sign to that of spin fluctuations 
dominant at higher temperatures.\cite{Narikiyo} 

  The author is grateful to Kenji Ishida 
for valuable discussions on various NMR experiments. 

\newpage

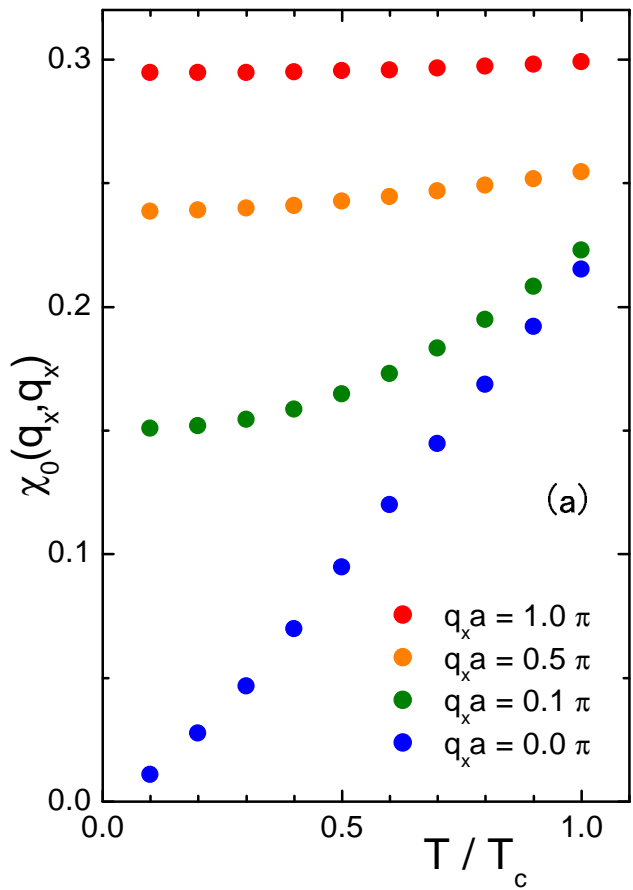
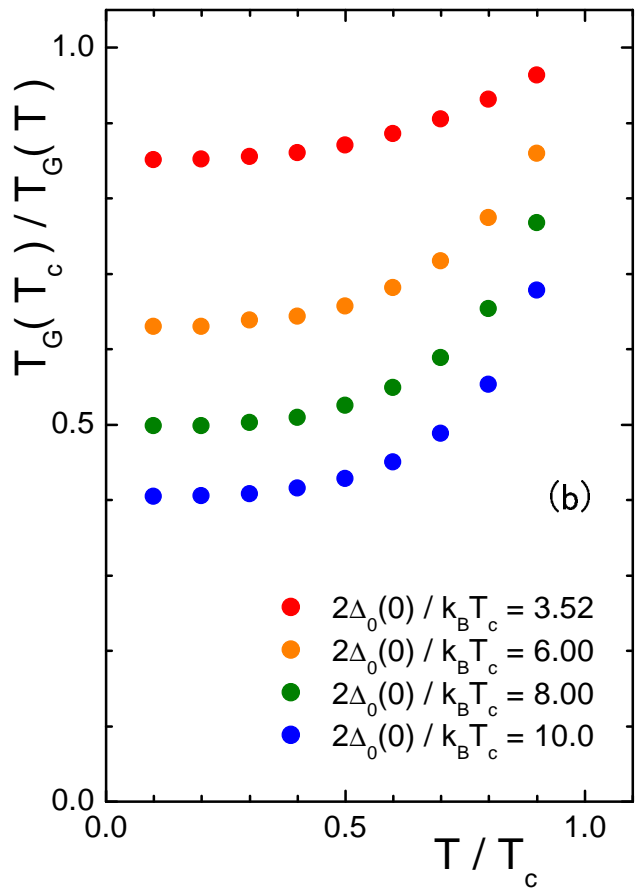
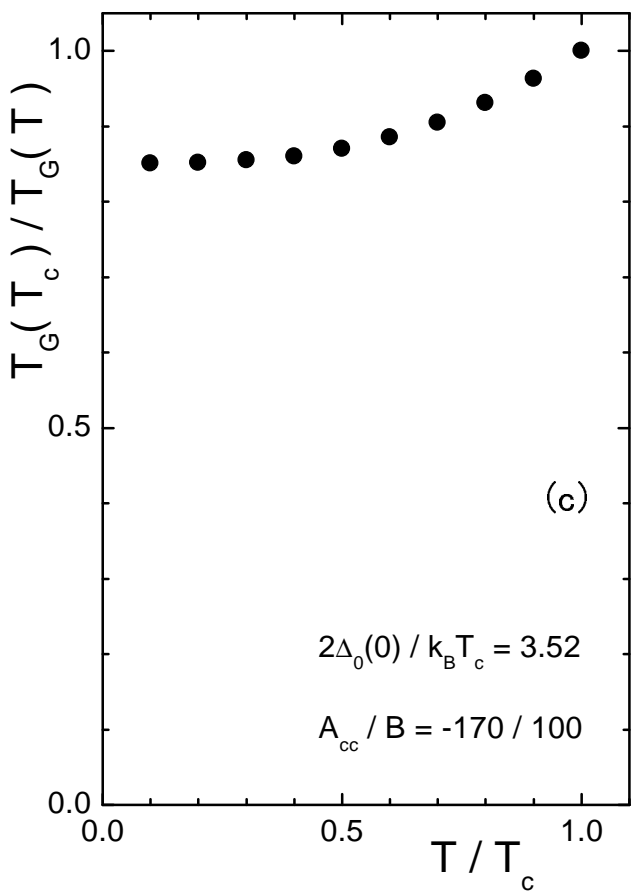
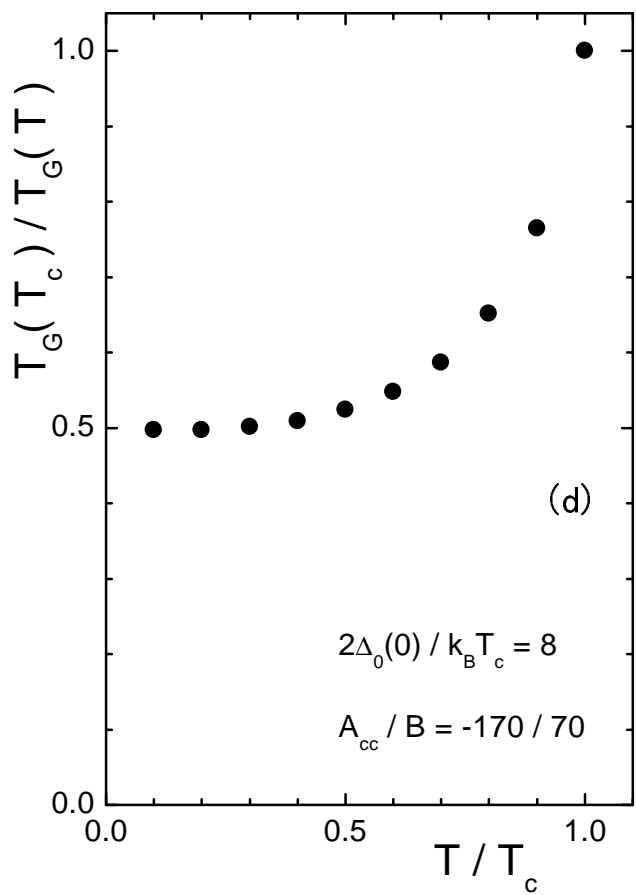
\end{document}